\documentclass[aps,prl,reprint,groupedaddress]{revtex4-1}
\usepackage{times}
\usepackage{chemarrow} % for xarrows 
\usepackage{amsmath,amssymb,mathrsfs}
\usepackage{graphicx,epsfig}
\usepackage{bbm}
\usepackage{extarrows,chemarrow,xypic} % 
\usepackage{hyperref}
\usepackage{microtype}
\usepackage{xcolor}
\DisableLigatures[f]{encoding = *, family = *}
\usepackage[none]{hyphenat}

\def\rd{{\rm d}}

\def\vw{{\bf w}}

\begin{document}

%\preprint{}

\title{Generalized Boltzmann distributions for systems strongly coupled to large finite bath -- a microcanonical approach}


\author{Yu-Chen Cheng$^1$}
\email[]{yuchench@u.washington.edu}
\author{Wenning Wang$^2$}
\email[]{wnwang@fudan.edu.cn}
\author{Zhiyue Lu$^3$}
\email[]{zhiyuelu@unc.edu}
\author{Hong Qian$^1$}
\email[]{hqian@u.washington.edu}
%\thanks{}
\affiliation{$^1$Department of Applied Mathematics, University of Washington, Seattle, WA 98195, U.S.A.\\
$^2$Department of Chemistry, Fudan University, Shanghai 200438, P.R.C.\\
$^3$ Department of Chemistry,
University of North Carolina-Chapel Hill,
Chapel Hill, NC 27599, U.S.A.}


%\date{\today}

\begin{abstract}

The theory of probability shows that, as the fraction $X_n/Y\to 0$, the conditional probability for $X_n$, given $X_n+Y \in h_{\delta}:=[h, h+\delta]$, has a limit law
$f_{X_n}(x)e^{-\psi_n(h_\delta)x}$, where $\psi_n(h_\delta) $ equals to $[\partial \ln P(Y \in y_\delta)/\partial y]_{y=h}$ plus an additional term, contributed from the correlation between $X_n$ and bath $Y$. By applying this limit law to an isolated composite system consisting of two strongly coupled parts, a system of interest and a large but finite bath, we derive the generalized Boltzmann distribution law for the system of interest in the exponential form of a redefined Hamiltonian and corrected Boltzmann temperature that reflects the modification due to strong system-bath coupling and the large but finite bath. 

%a corresponding generalized Boltzmann distribution law. In contrast to previous results derived with the help of an additional larger bath and a pre-determined temperature, our approach treats the composite system as an isolated system. It allows us to derive 

%This result is applicable to the system of arbitrary size, and also allows for strong system-bath interactions. 

%Furthermore, we discuss the thermodynamic relation of free energy, internal energy, and entropy when the system itself reaches the thermodynamic limit but the bath may or may not reach the ideal thermal-bath limit. 

%Since $\delta$ is infinitesimal, the theory shows a logic equivalence between the conditional probability setup and traditional energy conservation in dynamics. We obtain the state space density for a system strongly coupled to a heat bath. A discrete version of the theorem yields the grand canonical distribution without the Gibbs paradox.  In terms of certain invariant criteria with respect to $h_\delta$, the concept of a thermodynamic reservoir is given.

\end{abstract}

%\keywords{}

\maketitle

{\bf\em Introduction |} Gibbs' ensemble theory properly concerns with finite size systems that are in contact with a {\em temperature bath}.  It reproduces the classical thermodynamic results as the limiting behavior when the system size approaches infinity, e.g., {\em thermodynamic limit}. The Laplace transform of partition functions becomes the Legendre transform of thermodynamic potentials in the limit. The theory has been widely checked against experiments on finite and even small systems of liquids, solutions, and macromolecules. In biophysical chemistry, it has been used as a natural law for further understanding molecular interactions \cite{schellman}.

Through a rigorous mathematical analysis, we recently argued \cite{cheng-qian} that not only the macroscopic thermodynamics is an emergent phenomenon, the very Gibbs' formalism itself is a result of a limit law according to the theory of {\em conditional probability}.  The limit law is to the canonical distribution what the central limit theorem is to the Gaussian fluctuation theory developed by Einstein and Landau \cite{landau}.  And as in the theories of phase transition \cite{pwanderson} and the passing from quantum mechanics to quantum chemistry via Born-Oppenheimer approximation, the mathematical limit involves subtleties that have fundamental importance \cite{vulpiani-book}.

\begin{figure}[h]
\includegraphics[width=0.4\textwidth]{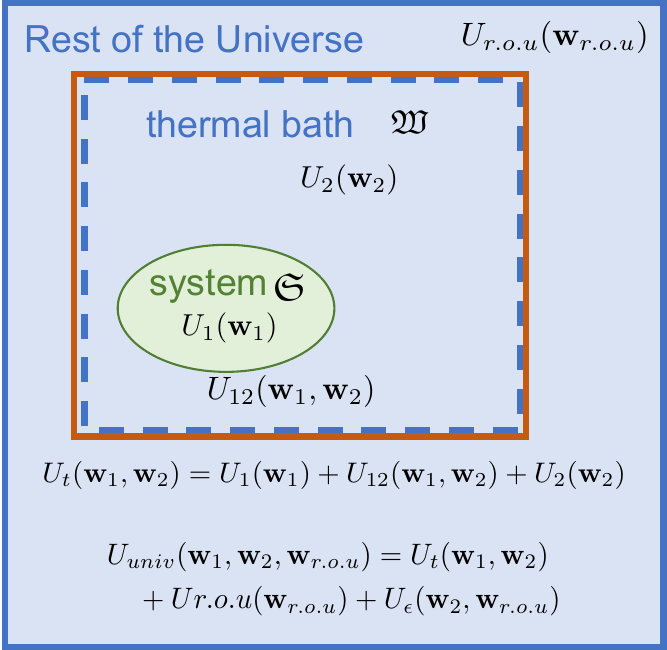}
\caption{
Illustrate are two approaches to derive the statistical law for a system coupled to a bath. Previous approaches assume that the composite system follows a canonical distribution $P(\vw_1,\vw_2)$ guaranteed by the Rest of the Universe (additional bath). In contrast, we propose an all-inclusive picture, where the system and bath are isolated (solid-line box). This approach assumes that the composite system follows from the microcanonical ensemble with a given conserved total energy, without presuming a temperature. It allows us to refine results by various choices of the bath's size when the bath's size is large but finite before achieving the equivalence of ensembles. 
}

\centering
\label{fig-1}
\end{figure}

The aim of this paper is to generalize the Gibbs' theory to systems that are either weakly or strongly coupled to the surrounding bath, and the bath can be finitely large rather than an infinitely large ideal thermal bath. Let us consider a system $\mathfrak{S}$ coupled to a surrounding bath $\mathfrak{W}$ with a  setup:  (i) the composite system $\mathfrak{S} + \mathfrak{W}$ can be  {\em heterogeneous} as the physical or chemical identity of $\mathfrak{S}$ can be distinct from $\mathfrak{W}$, e.g., a RNA molecule immersed in aqueous solution \cite{jarzynski}; (ii) interactions between $\mathfrak{S}$ and $\mathfrak{W}$  can be non-negligible. Then our goal is to describe the distribution law of the state of $\mathfrak{S}$ from an all-inclusive perspective by treating the composite system $\mathfrak{S} + \mathfrak{W}$ as an isolated system entirely independent of the outside. See FIG. \ref{fig-1}.

{\bf\em Previous works |} All-inclusive picture of an isolated composite system consisting of a system and a bath has been widely used in deriving thermodynamics laws and statistical mechanics relations.  Previous work \cite{hilbert2014thermodynamic} has obtained the {\em thermodynamic laws} of system $\mathfrak{S}$ that is part of an isolated composite system $\mathfrak{S} + \mathfrak{W}$. In this letter, we also take the all-inclusive picture of $\mathfrak{S} + \mathfrak{W}$ but with a focus on obtaining the statistical law for $\mathfrak{S}$, where the system can be strongly coupled to a {\em large but finite} bath. Our statistical law is obtained in a scenario between the two extremes: In \cite{hanggi2016meaning}, when the bath is infinitely large and weakly coupled to system $\mathfrak{S}$, the state of $\mathfrak{S}$ simply follows the classical Boltzmann distribution based on the classical canonical ensemble theory. On the other hand in \cite{khinchin, campisi2009finite}, when the bath size is finite and comparable to the system size, even if the system couples weakly to the bath, the distribution of $\mathfrak{S}$ can not be written in a universal exponential form (no generalized Boltzmann form). 

Beyond previous work, for a large but finite bath, one can anticipate that the statistical law of its contacted system $\mathfrak{S}$ may follow a Boltzmann-like form with proper corrections. A large but finite bath is very common in molecular dynamics (MD) simulations where the computational cost grows exponentially with the number of atoms of the total composite system. The all-inclusive picture of isolated composite system $\mathfrak{S} + \mathfrak{W}$ is a perfect description of a microcanonical MD simulation inside a finite box. Inside the simulation box system of interest $\mathfrak{S}$ can be a macromolecule or a molecular complex immersed in surrounding explicit solvent molecules $\mathfrak{W}$. Due to the non-negligible interaction between the system and the solvent, one can expect that the internal energy of $\mathfrak{S}$ needs to be redefined to take into account the interaction energy and the {\em Boltzmann temperature} of $\mathfrak{S} + \mathfrak{W}$ may also require modification.

{\bf\em Conditional probability theory |} Based on the above setup, current mathematical theories,  {\em the Gibbs conditioning principle}  \cite{Zabell, vC-C,stroock, dembo}  and  {\em the equivalence of ensembles} \cite{khinchin, martin, Deuschel, lewis, Georgii, touchette,Friedli}, are not sufficient to describe the statistical law of $\mathfrak{S}$ for the following reasons: (i) the former theory is based on the mathematical framework of $\mathfrak{W}$ as a large number of identical copies of $\mathfrak{S}$, i.e. {\em homogeneous} composite systems. (ii) the later theory requires bath $ \mathfrak{W}$ being infinitely large as a necessary condition for the equivalence of ensembles.

%These theories require $\mathfrak{W}$ as a large number of identical copies of $\mathfrak{S}$, i.e. homogeneous composite systems; furthermore, they fail in certain circumstance when the system has critical phenomena \cite{martin, touchette, Friedli}. 

To solve above issues, a new approach based on the conditional probability theory \cite{cheng-qian} is presented in this work: Let $U_1$ and $U_2$ be the bare energy of $\mathfrak{S}$ and  $\mathfrak{W}$, respectively, and $ U_{12}$ be their interaction energy. In this letter, we focus on two cases: (i) if $ U_{12} \ll U_1 $,  the composite system $\mathfrak{S} + \mathfrak{W}$ is called a {\em weakly coupled system} and (ii) if $U_1 \simeq U_{12}$,  the composite system $\mathfrak{S} + \mathfrak{W}$ is called a {\em strongly coupled system}. In case (ii), we construct a pseudo-Hamiltonian $\hat{U}_1$ for  system $\mathfrak{S}$  so that the microstates of the system are equally probable inside each pseudo-energy shell defined by $\hat{U}_1$. In case (i), by the {\em principle of equal a priori probabilities} \cite{huang}, $\hat{U}_1$ coincides with the bare Hamiltonian $U_1$. Given a conserved total energy $U_t = U_1+U_{12}+U_2$, we can define the conditional probability distribution of the pseudo-energy, $P_{\hat{U}_1|U_t}$. 

%This microcanonical treatment of isolated composite system $\mathfrak{S} + \mathfrak{W}$ which is strongly coupled allows us to derive and generalize strongly coupling thermodynamic laws that were previously obtained with the help of an additional larger bath \cite{jarzynski}. 

%It is worth to note that our microcanonical approach allows us to discern two strongly coupled scenarios: (1) strong coupling w.r.t system of interest, and (2) strong coupling w.r.t. both system and bath.     

An important result in this letter shows that as the size of bath $\mathfrak{W}$ is relatively large to system $\mathfrak{S}$, then $\hat{U}_1 \ll U_t$ leads to an emergent law: $P_{\hat{U}_1|U_t}$ is asymptotic to $P_{\hat{U}_1}$ weighted by an exponential factor whose exponent is determined by both of fluctuations of $U_t - \hat{U}_1$ and a correlation between $\hat{U}_1 $ and $U_t - \hat{U}_1$. This emergent law for $P_{\hat{U}_1|U_t}$ is based on a key theorem provided later. Furthermore, since the way of construction of  pseudo-Hamiltonian $\hat{U}_1$ guarantees all microstates are equally probable in each pseudo-energy shell, the state space density of $\mathfrak{S}$ turns out to be a purely exponential distribution in terms of the pseudo-Hamiltonian with a modified thermodynamic parameter (inverse of temperature). This novel result of the state space density is named as the {\em generalized Boltzmann distribution law} in this letter.

%{\em Outline of the letter |}  We start with a statement of the theorem on conditional probability with interpretations. Then we apply it to two scenarios separately: (i) $U_{12} \ll U_1 \ll U_2$  (weakly coupled systems) and (ii) $U_1 \simeq U_{12}\ll U_2$ (strongly coupled systems). Our approach based on the microcanonical assumption of the composite system  $\mathfrak{S} + \mathfrak{W}$ is in contradistinction to the canonical assumption of the composite system \cite{jarzynski,talkner}, which requires a concept of a larger bath, e.g., the rest of universe. See FIG. \ref{fig-1}. 

{\bf\em Theorem, a limit law for a sequence of conditional probabilities |} We here state the key theorem without the proofs, see \cite{cheng-qian} for the full mathematical details.  Consider a sequence of non-negative random variable $X_n\in\mathbb{R}$ and another non-negative random variable $Y\in\mathbb{R}$ which needs not to be independent of $X_n$, on a probability space $(\Omega,\mathcal{F},P)$.  Denoting $H_n=X_n+Y$ and  $X_n / Y \rightarrow 0$ as $n\to\infty$ \cite{foot.note.an}. Then the sequence of conditional probability density functions has an asymptotic expression \cite{foot.note.kl}:  
\begin{subequations}
\begin{equation} \label{1}
   f_{X_n|H_n}(x;h_\delta ) \simeq       
    Z^{-1}_n(h_\delta)f_{X_n}(x) e^{-\psi_n(h_\delta)x},
\end{equation}
as $n\to\infty$, where $h_\delta := [h, h+\delta]\in\mathbb{R}$, $f_{X_n}(x)$ is the marginal distribution of the correlated pair $(X_n,Y)$, $Z_n$ is the normalization factor. Furthermore,  the parameter on the exponent follows the formula
\begin{align}
    \psi_n(h_\delta) &= 
  \frac{\partial\ln P(Y \in y_\delta) }{\partial y}
  \bigg\rvert_{y=h} \label{temperature-0}\\
  +& \left[\frac{\partial \ln C_n(y_\delta ; x) }{\partial y} - \frac{\partial \ln C_n(y_\delta ; x) }{\partial x}\right]_{x=0,y=h}, \label{temperature}
\end{align}
where $y_\delta := [y, y+\delta]$ and
\begin{align}
    C_n(y_\delta ; x) = \frac{P(Y\in y_\delta | X_n =x)}{P(Y\in y_\delta)}.
\label{temperature-2}
\end{align}
\end{subequations}
%If $Y$ is independent from $X_n$, $C_n=1$ and $\psi_n(h_\delta)$ is independent of $n$; then the inverse temperature is solely determined by the property of the bath $P(Y \in [y, y+\delta])$ near $y=h$. 

%In the following,  $(\partial \ln P(Y \in y_\delta)/\partial y)_{y=h} $ is denoted  by $\beta^*(h_\delta)$.

The theorem can be interpreted with the following physical assumptions:  First, we denote the system energy $X_n$ indexed by $n$, where $n$ tracks the relative size of the bath to the system.  In the limit of $n \rightarrow \infty$,  $X_n/Y$ goes to zero and the bath becomes an {\em ideal thermal-bath}. Second, the total mechanical energy of the isolated composite system $X+Y=h$ is conserved, subjected to infinitesimal uncertainty $\delta$. This corresponds to the assumption that the total system is isolated from the rest of the universe, i.e., sampled from a {\em microcanonical ensemble} of total energy $[h, h+\delta]$.  Third, when the strong coupling is negligible compared to the bath's energy in the ideal thermal-bath limit, a universal thermodynamic parameter (i.e., inverse temperature) emerges, $\beta^*(h_\delta) :=\partial \ln P(Y \in h_\delta)/\partial y$ which is uniquely determined by the fluctuations of the bath. This is reflected in the correlation-related term \eqref{temperature}  being negligible in comparison to $\beta^*$ in \eqref{temperature-0}  as $n \rightarrow \infty$.

\begin{figure}[h]
\includegraphics[width=0.32\textwidth]{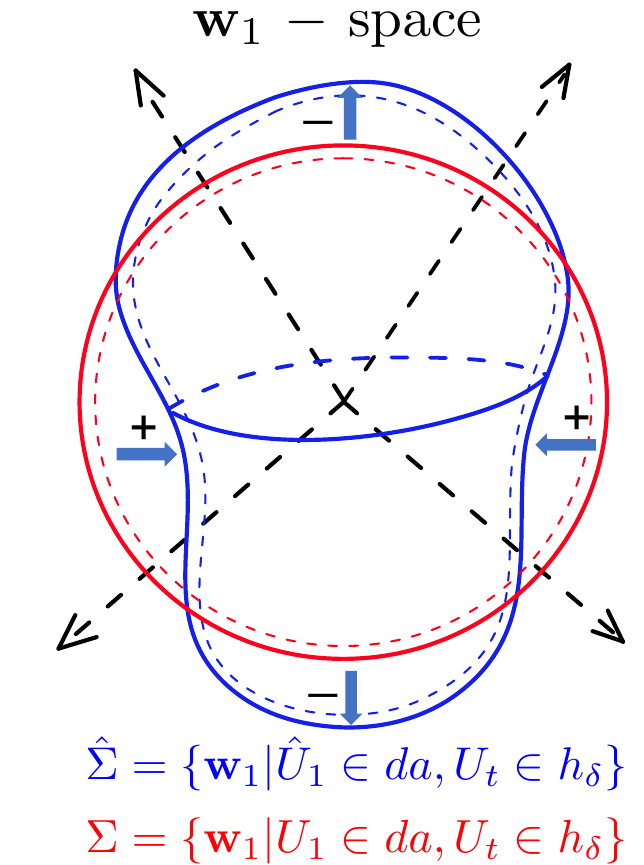}
\caption{
Sketched here in the $6n_1$-dimensional phase space for the system is the energy shell $\Sigma$ (red) and the pseudo-energy shell $\hat \Sigma$ (blue). For weakly coupled systems, the energy shell and the pseudo-energy shell are identical, and the probability density of $\vw_1$ is uniform in the shell. However, when the interaction between system and bath becomes non-negligible, $\hat \Sigma$ and $\Sigma$ are no longer identical, and the equal probability shell is defined by $\hat \Sigma$ rather than $\Sigma$. The arrow shapes with ``$+/-$" signs refers to the deformation of $\hat \Sigma$ from $\Sigma$ due to the repulsive/attractive interaction between system and bath.
}

\centering
\label{fig-2}
\end{figure}

%in particular, if $Y$ is independent from $X_n$, $C_n=1$, then a universal thermodynamic parameter (i.e., inverse temperature) arise, $\beta^*(h_\delta) :=\partial \ln P(Y \in h_\delta)/\partial y$ which is uniquely determined by the fluctuations of the heat bath. We will show that this independence corresponds to the assumption $U_{12} \ll U_1$, i.e. weakly coupled systems.  

%if the system of interest is asymptotically uncorrelated with the bath in the ideal thermal-bath limit, then a universal thermodynamic parameter (i.e., inverse temperature) arise, $\beta^*(h_\delta) :=\partial \ln P(Y \in h_\delta)/\partial y$ which is uniquely determined by the fluctuations of the heat bath. This is reflected in the correlation-related term \eqref{temperature} converging to $0$ as $n \rightarrow \infty$. 

%In the following lemma, $X$ is indexed by $n$ to represent an infinitesimal system and no assumption of independence between ``system of interest'' $X$ and ``heat bath'' $Y$ is made.  Thm 1. shows, however, the importance of $X,Y$ independence in defining a unambiguous equilibrium temperature for all $X$s that are in contact with a given $Y$.  Thm 2. is an extension of the lemma to integer random variables.  Thm 3. further clarifies the condition under which the temperature is uniquely defined for a heat bath irrespective of the details of the ``contact''.

%New result of applying the Lemma to strongly coupled systems is presented in a later section. 

%\begin{eqnarray}
   %Z_n(h_\delta) &=& \int_0^{h+\delta}  f_{X_n}(x) e^{-\psi_n(h_\delta)x}\rd x.
%\end{eqnarray} 

{\bf\em  Weakly coupled heterogeneous systems  |} 
Let us first apply the theorem to weakly coupled heterogeneous systems and assume that  system $\mathfrak{S}$ contains $n_1$ particles and  bath $\mathfrak{W}$ contains $n_2$ particles and that their interaction is weak. Consider a probability space $(\Omega,\mathcal{F},P)$ that $\Omega\subset\mathbb{R}^{6n_1+6n_2}$ with state  $(\mathbf{w}_1,\mathbf{w}_2)$, $\vw_1 \in\mathbb{R}^{6n_1}$, $\vw_2 \in\mathbb{R}^{6n_2}$.  Furthermore $\vw_1$ and $\vw_2$ are independent w.r.t. $P$. In this application, $\Omega$ represents the compact, continuous phase space of the composite system $\mathfrak{S} + \mathfrak{W}$, and $\vw_1$ is the microstate variable for system $\mathfrak{S}$ and $\vw_2$ is the microstate variable for bath $\mathfrak{W}$, respectively; $\mathcal{F}$ stands for the sigma-algebra of $\Omega$, i.e. all possible sets of outcomes in the phase space; and $P$ is a prior probability measure on  $\mathcal{F}$.

For weakly coupled systems, mechanical energy functions satisfy that $U_t(\vw_1, \vw_2)=U_1(\vw_1)+ U_{12}(\vw_1, \vw_2)+U_2(\vw_2)$ and $U_{12}(\vw_1, \vw_2) \ll U_1(\vw_1) \ll U_2(\vw_2)$. By denoting $X_n := U_1(\vw_1)$ and $Y:= U_2(\vw_2) + U_{12}(\vw_1, \vw_2) $, the limit distribution in \eqref{1} implies that the energy distribution of $\mathfrak{S}$ contained given a conserved energy $U_t$ of $\mathfrak{S}+ \mathfrak{W}$ follows 
\begin{align}
     \label{canonical-gibbs-measure}
  & P(U_1(\vw_1)  \in \rd a | U_t(\vw_1,\vw_2) \in h_\delta) \nonumber \\ 
   \simeq \ & Z^{-1}(\beta^*) e^{-\beta^*a} \times P(U_1(\vw_1)  \in \rd a ),
\end{align}
in which $\rd a$ is a short notation for  system's energy shell $[a, a+ \rd a]$. It is worth to note that the thermodynamic parameter $\beta^*$ of the exponential function in \eqref{canonical-gibbs-measure} is uniquely determined by the energy fluctuations of the bath. This is a result of $C_n = 1$ for all $n$ in  \eqref{temperature-2} due to negligible $U_{12}$ in comparison with both $U_1$ and $U_2$, which is corresponding to the assumption of additive functions for $U_1, U_2$ in classical statistical mechanics.

In order to further justify the distribution law for microstate variable $\vw_1$ of $\mathfrak{S}$, we need the following fact
\begin{align} \label{equ.priori-1}
    &f(\vw_1 | U_1(\vw_1) \in \rd a,  U_t\in h_\delta) \nonumber \\
    =&  f(\vw_1 | U_1(\vw_1) \in \rd a) \propto  \frac{\mathbf{1}_{\{U_1(\vw_1) \in \rd a \}}}{P(U_1(\vw_1) \in \rd a )}, 
\end{align}
where the first equation is due to independent $(U_1, U_2)$ and $f(\vw_1 | U_1(\vw_1) \in \rd a) $ has to be uniformly distributed in energy shell $\rd a$ by the assumption of the principle of equal a priori probabilities. Then by integrating $P_{U_1 | U_t}(\rd a; h_\delta)$ in \eqref{canonical-gibbs-measure} with $f(\vw_1 | U_1 \in \rd a,  U_t\in h_\delta)$ in \eqref{equ.priori-1}  over all $a$, we can show  
\begin{equation} \label{canonical-gibbs}
  f(\vw_1| U_t \in h_\delta ) \simeq Z^{-1}(\beta^*) e^{-\beta^*U_1(\vw_1)},
\end{equation}
which justifies the classical Boltzmann distribution law for the weakly coupled systems.

%In this application, $\Omega$ represents the compact, continuous phase space of the total system with microstate variable $(\vw_1,\vw_2) $. Let $\Omega_1$ be the phase space for $\vw_1$ and $\Omega_2$ be the phase space for $\vw_2$. Then the mathematical structure of our theory for weakly coupled heterogeneous systems is as follows:
%\begin{subequations}
%\begin{align}
%    & ( \Omega , \mathcal{B}(\Omega), P) \xrightarrow{\pi_1} (\Omega_1 , \mathcal{B}(\Omega_1 ), \mu_1) \xrightarrow{U_1} (\mathbb{R} , \mathcal{B}(\mathbb{R} ), F_{U_1}),  \nonumber\\
%    & ( \Omega , \mathcal{B}(\Omega), P) \xrightarrow{\pi_2} (\Omega_2 , \mathcal{B}(\Omega_2 ), \mu_2) \xrightarrow{U_2} (\mathbb{R} , \mathcal{B}(\mathbb{R} ), F_{U_2}),\nonumber
%\end{align}
%\end{subequations}
%in which $\pi_1, \pi_2$ are the projection maps with the corresponding induced measures $\mu_1, \mu_2$, and $U_1, U_2$ are the functions for observables with the distributions $F_{U_1}, F_{U_2}$. To begin with the theorem, we have $F_{U_1 | U_t}$ at the level of observables; furthermore, by the principle of equal a priori probabilities for all $\vw_1$ in each thin shell of $U_1$, we then obtain $f(\vw_1 | U_t \in h_\delta)$ at the level of microstates. This can be justified by 
%(using $\rd a := [a, a+\rd a]$ for simplicity)

The strength of our approach can be seen when it is applied to heterogeneous systems: Let us consider the example for a macromolecule immersed in an aqueous solution with negligible interactions. Now, $\vw_1 $ represents the microstate of the molecule and $\vw_2 $ represents the microstate of the solution. In this case, since the molecule is distinct to the aqueous solution, so a justification of the distribution of $\vw_1$  was missing in the previous mathematical theories based on homogeneous composite systems \cite{Zabell, vC-C,stroock, dembo}. By our approach, one only need to 
satisfy $U_{12}(\vw_1, \vw_2) \ll U_1(\vw_1) \ll U_2(\vw_2)$ to obtain the energy distribution of $U_1(\vw_1)$ in \eqref{canonical-gibbs-measure}; Furthermore,  given the principle of equal a priori probabilities for $\vw_1$ \cite{foot.note.molecule}, the classical Boltzmann distribution of $\vw_1$ in \eqref{canonical-gibbs} can be justified.

{\bf\em Strongly coupled  heterogeneous  systems | }  Here we further apply the theorem to strongly coupled heterogeneous systems, in which total energy is the sum of three non-negative functions
\begin{align}
    U_t(\vw_1,\vw_2)=U_1(\vw_1)+ U_{12}(\vw_1, \vw_2) +U_2(\vw_2),
\end{align}
and $U_{12}(\vw_1, \vw_2) $ is a non-negligible interaction energy in comparison with the order of $U_1(\vw_1)$. Therefore, in contrast to weakly coupled systems, if we simply group $U_{12}(\vw_1, \vw_2)$ and $U_2(\vw_2) $ together into a new observable  $Y:= U_{12}(\vw_1, \vw_2) +U_2(\vw_2)$, then $X_n := U_1(\vw_1)$ and $Y$ are no longer independent. Since the theorem does not require an independent pair of random variables, parallel to the case of weakly coupled systems, we still have an asymptotic expression of $P_{U_1 | U_t}$. However, since $(X_n,Y)$ is non-independent, the first equation in \eqref{equ.priori-1} is no longer true. In other words, microstates $\vw_1$ might not be equally probable in each energy shell defined by bare Hamiltonian $U_1$ due to the strong coupling with the bath, and thus we are not be able to further derive the classical Boltzmann distribution law for $\mathfrak{S}$ in terms of $U_1$.

Alternatively, we can reconstruct energy shells such that  $\vw_1$ is uniformly distributed in each of them. Let us consider  a ``pseudo-Hamiltonian" $\hat{U}_1(\vw_1)$ satisfying
\begin{align} \label{equ.priori}
    f(\vw_1 | \hat{U}_1(\vw_1) \in \rd a,  U_t\in h_\delta) \propto  \frac{\mathbf{1}_{\{\hat{U}_1(\vw_1) \in \rd a \}}}{P(\hat{U}_1(\vw_1) \in \rd a )}.
\end{align}
The existence of this pseudo-Hamiltonian is discussed later. Now, despite that the redefined pair of observables $\hat{X}_n:=\hat{U}_1(\vw_1)$ and $\hat{Y}:= U_t(\vw_1, \vw_2) - \hat{U}_1(\vw_1)$ are still dependent on each other. Since the way to construct the pseudo-energy by \eqref{equ.priori}  guarantees that $\vw_1$ is uniformly distributed in each pseudo-energy shell $\{\vw_1 | \hat{U}_1(\vw_1) \in \rd a, U_t\in h_\delta\}$,  even when $\vw_1$ has strong interactions with the bath, we can further obtain a generalized Boltzmann distribution law
\begin{align}
\label{strong-coupled-density}
   f^{(m)}(\vw_1 | U_t \in  h_\delta ) \propto e^{- \psi(h_\delta) \hat{U}_1(\vw_1) },
\end{align}
in which the superscript $(m)$ indicates that we assumes that $\mathfrak{S} + \mathfrak{W}$ is an isolated total system with conserved total energy (microcanonical ensemble treatment). See Fig. \ref{fig-2} for the construct of the pseudo-energy shells.

 It is worth to remind that thermodynamic parameter $ \psi(h_\delta)$  of the exponential function in \eqref{strong-coupled-density} is given by the equation
\begin{subequations}
\begin{align} 
 \psi_n(h_\delta) &= 
  \frac{\partial\ln P(\hat{Y} \in y_\delta) }{\partial y}
  \bigg\rvert_{y=h} \label{corrected-temp-0}\\
  &+ \left[\frac{\partial \ln C_n(y_\delta ; x) }{\partial y} - \frac{\partial \ln C_n(y_\delta ; x) }{\partial x}\right]_{x=0,y=h}.  \label{corrected-temp-1}
\end{align}
As $\delta \rightarrow 0$,  function $C(y,x)$ is known as the {\em copula density}
\begin{align}  \label{corrected-temp-2}
    C_n(y,x) = \frac{f_{\hat{Y}|\hat{X}_n}(y;x)}{f_{\hat{Y}}(y)} = \frac{f_{\hat{Y},\hat{X}_n}(y,x)}{f_{\hat{Y}}(y)f_{\hat{X}_n}(x)},
\end{align}
and it is intimately related to the {\em mutual information}
\begin{align} \label{corrected-temp-3}
    I(\hat{Y}; \hat{X}_n) = \int_y \int_x f_{\hat{Y},\hat{X}_n}(y,x) \ln \frac{f_{\hat{Y},\hat{X}_n}(y,x)}{f_{\hat{Y}}(y)f_{\hat{X}_n}(x)}.
\end{align}
\end{subequations}
 Eqs. \eqref{corrected-temp-0} - \eqref{corrected-temp-3} shed light on how to properly correct the temperature of strongly coupled systems when bath $\mathfrak{W}$ is large but finite. By \eqref{corrected-temp-0} and \eqref{corrected-temp-1}, the thermodynamic parameter of the generalized Boltzmann distribution, i.e. inverse of temperature, depends on the index $n$ that indicates the choice of the relative size of bath $\mathfrak{W}$ to the size of system $\mathfrak{S}$. As $n$ is finite, on top of $\beta^*$,  $\psi_n(h_\delta)$ has to be corrected by an additional term caused by the correlation function \eqref{corrected-temp-2}, and it is a consequence of $U_{12} \simeq U_1$ for strongly coupled systems. On the other hand, since the strongly coupled systems considered in this letter only involve {\em short range interactions},  as the bath size increases, the interaction energy is negligible with respect to the bath's energy by the orders $U_1 \simeq U_{12}\ll U_2$. Therefore, when bath $\mathfrak{W}$ achieves the ideal thermal-bath limit as $n \rightarrow \infty$, there should emerge a limit of the temperature uniquely determined by the bath, i.e. $ \lim_{n \rightarrow \infty}\psi_n(h_\delta) = \beta^*$.

Recently,  Jarzynski \cite{jarzynski}, Talkner and H\"anggi \cite{talkner2016open,talkner2020comment,talkner}, have discussed thermodynamic laws of systems that are strongly coupled to a big bath. They have taken an assumption that the total system of the system plus bath is in contact with a bigger background bath (or the rest of the universe). See FIG. \ref{fig-1}. This allows for the total system to be canonically distributed: $f^{(c)}(\vw_1,\vw_2) \propto \exp\{-\kappa\left[ U_1(\vw_1) + U_{12}(\vw_1, \vw_2)+  U_{2}(\vw_2)\right]\} $. Here we use the superscript $(c)$ to denote the probability under the assumption of canonically distributed total system. The marginal density then
\begin{align}
\label{strong-coupled-density-2}
  &f^{(c)}(\vw_1) \propto e^{-\kappa\left[ U_1(\vw_1) + \phi^{(c)}(\vw_1;\kappa)\right]},\\
   &\phi^{(c)}(\vw_1;\kappa) =  -\frac{1}{\kappa}\ln \frac{ \int \rd \vw_2 e^{-\kappa\left[U_{12}(\vw_1, \vw_2) + U_{2}(\vw_2)\right]} }{ \int \rd \vw_2 e^{-\kappa U_{2}(\vw_2)} }, \nonumber
\end{align}
where $U_1(\vw_1) + \phi^{(c)}(\vw_1;\kappa)$ is often called the {\em Hamiltonian of mean force}; As the function depends only on the positions, it
matches Kirkwood's potential of mean force \cite{kirkwood}; the force is acting on 
$\vw_1$ and averaged over all the 
$\vw_2 \in\mathbb{R}^{2n_2}$.

 As $n \rightarrow \infty$, i.e. the ideal thermal-bath limit, the theory of equivalence of ensembles \cite{martin, Deuschel, lewis, Georgii} between the microcanonical treatment and the canonical treatment of the composite system  $\mathfrak{S}+\mathfrak{W}$  is applicable and thus $ f^{(m)} = f^{(c)}$ in Eqs \eqref{strong-coupled-density} and \eqref{strong-coupled-density-2} with the same temperature, i.e.,  $\lim_{n \rightarrow \infty}\psi_n(h_\delta) = \beta^* =  \kappa$. The above equivalence leads to an essential definition of our pseudo-Hamiltonian
\begin{equation}
\begin{aligned} \label{pair.eq}
     \hat{U}_1(\vw_1) = U_1(\vw_1) + \phi^{(c)}(\vw_1;\kappa),
\end{aligned}
\end{equation}
which has a novel and significant physical interpretation: The pseudo-Hamiltonian $\hat{U}_1(\vw_1)$ given in  \eqref{equ.priori}  guarantees equally probable $\vw_1$ in each energy shell defined by $\hat{U}_1(\vw_1)$, and Eq. \eqref{pair.eq} indicates that this pseudo-Hamiltonian actually exists and coincides with the Hamiltonian of mean force in the ideal thermal-bath limit. This key result reveals the reason why the separation of degrees of freedom between the system and its bath relies on the Hamiltonian of mean force in the canonical treatment of strongly coupled thermodynamics. We shall emphasize that the choice of Hamiltonian of mean force to redefine the Hamiltonian of system $\mathfrak{S}$ has a pivital meaning but was not mentioned previously - it guarantees that microstate variable $\vw_1$ is uniformly distributed in each pseudo-energy shell so that the generalized Boltzmann distribution law is valid to describe the phase space density of  system $\mathfrak{S}$ in terms of this redefined Hamiltonian.

{\bf \em Legendre transform in thermodynamics |} With one additional assumption, the result in Eq. \eqref{1} gives rise to the Legendre transform in thermodynamics: In the thermodynamic limit of system  $\mathfrak{S}$, with its thermodynamic energy and entropy being ``extensive quantities'' \cite{foot.note.ldp}, while keeping the ratio of the bath size to the system size finite, the logarithm of the  normalization factor in \eqref{1} gives rise to a thermodynamic potential in the unit of $(\psi_n)^{-1}$ 
\begin{subequations}
\begin{eqnarray}
	F(\psi_n) &\equiv& -(\psi_n)^{-1}\ln Z  \ = \  
          -(\psi_n)^{-1}\ln \int_0^{h+\delta} e^{S(a) -\psi_n x }\rd x
\nonumber\\
	 &\simeq& \Big[E -(\psi_n)^{-1} S(E)\Big]_{\rd S(E)/\rd E= \psi_n }.
\label{eq-5}
\end{eqnarray}
which resembles the result of the classical thermodynamics. However, due to the non-negligible strong interaction between system and the bath, we  find the thermodynamic relation with a modified inverse temperature $\psi_n$. To fully recover the classical thermodynamics result, we can take the ideal bath limit where bath's size is infinitely large compared to the system and the system-bath interaction (i.e. $U_1 \simeq U_{12}\ll U_2$). In this ideal bath limit,  we can recover the classical thermodynamics result:
\begin{align} \label{thermodynamics}
    	F(\beta^*) &\simeq\Big [E -(\beta^*)^{-1} S(E)\Big]_{\rd S(E)/\rd E = \beta^* },
\end{align}
\end{subequations}
where inverse of $\beta^*=\partial \ln P(Y \in h_\delta)/\partial y$ is exactly the classical thermodynamic temperature.

\section{Acknowledgement} 

We thank Michael V. Berry, Chris Burdzy, Hao Ge, Yao Li,  Matt Lorig,  
%Lamberto Rondoni, 
Lowell Thompson, Hugo Touchette, Angelo Vulpiani, Peter  H\"anggi, %Xiangjun Xing, 
and Ying-Jen Yang for discussions and comments.
%\end{acknowledgments}



\begin{thebibliography}{00}
\bibitem{foot.note.an}
Since $X_n$ and $Y$ are both random variables, this ratio is rigorously defined by $\mathbb{E}[X^2_n]/\mathbb{E}[Y^2] \rightarrow 0$.
\bibitem{foot.note.kl}
The two probability density functions on lhs and rhs, in term of their KL divergence, has the order of $O(a_n)$, where $a_n = \mathbb{E}[X^2_n]/\mathbb{E}[Y^2]$.
\bibitem{foot.note.molecule}
If all the potential and kinetic energy of the macromolecule are considered, then this is a mechanical system and  thus the  principle of equal a priori probabilities is guaranteed.
\bibitem{foot.note.ldp}
In the thermodynami limit, energy $x$  and entropy $S(x)\equiv \ln f_X(x)$ are both proportional to the system's size $\propto\ V$, where $V\to\infty$ represents  the size of   system $\mathfrak{S}$ and the number of particles in $\mathfrak{S}$ both approaching infinity while keeping a constant density.
This is closely related to the large deviations principle in the theory of probability, and the widely cited ``maximum term method'' in standard textbook.
\end{thebibliography}
\end{document}